\documentclass[prl,twocolumn,showpacs]{revtex4}
\begin{document}
\title{Proteomic waves in networks of transcriptional regulators}
\author{A. S. Carstea}
\affiliation{\it Horia Hulubei National Institute
for Physics and Nuclear Engineering (IFIN-HH), Dept. of Theoretical Physics
407 Atomistilor, Magurele - Bucharest, 077125, Romania}

\begin{abstract}
A chain of connected genes with activation-repression links is analysed. 
It is shown that for various promoter activity functions (parametrised by Hill coefficient) 
the equations describing the 
concentrations of transcription factors are perturbed completely integrable 
differential-difference of KdV-type. In the case of large Hill coefficient the proteomic signal along the gene network is given by a superposition of perturbed dark solitons of defocusing 
differential-difference mKdV equation.
\end{abstract}
\maketitle
Genetically precoded responses of bacteria (and all living organisms) to external 
perturbations and signals is achieved through networks of genes of high complexity.
The interaction of genes aims at regulating each others' activity and thus leads to the specialised
response \cite{1}. Typically a gene is subject to the regulatory effect of  other genes which
can act on it in either an activating or a suppressing way, depending on the situation. The
predominant topic of many experimental and theoretical studies on genetic circuits so far
has been the combinatorial control of transcription, which, to a large extent
determines the connectivity of the network \cite{2}. It is thus very important to study 
and understand the dynamics of the gene regulatory networks.

This also will allow to design specific
gene regulatory networks performing certain functions. The problem of designing 
is quite complicated due to the poorly understanding of the "design principles". 
Up to now a large amount of studies have been performed {\it in vitro} and {\it in vivo} 
to understand the molecular mechanisms. As an outcome, at least at the bacterial 
level, it has been shown a hierarchical organization in 
motifs, modules and games \cite{3} - so we have a rather modular than a molecular organisation.

For example these networks display bistability, oscillatory behaviour
of some combinations of repressors and the level of description similar to electronic engineering blueprints seems to be quite appropriate\cite{4}.

In this paper we are going to show that under very simple assumptions a 
chain of transcriptional regulators having inducer-repressor design can display complex and in the 
same time controlable dynamics. Namely the "gene expression" dynamics of the gene can "propagate" to the 
other genes in a "solitary wave" manner and as a readout the concentrations of 
transcription factor proteins can have "solitonic" dependence with respect to the label of genes.
More precisely the proteic concentrations will obey, for various values of Hill 
coefficient in the promoter activity, differential-difference (semidiscrete) 
Korteweg de Vries (KdV) equation-types perturbed, of course, with terms containing 
degradation rates. We will discuss the non-Hill case as well. Because in a previous paper we analysed a general chain of genes having only activator-activator or repressor-repressor interaction and found conditions for the existence of bistable states \cite{5} it is natural to analyse also the alternating gene networks . The activity of a gene is regulated by other genes through the production of transcription
factor (TF) proteins. Physically, this is accomplished through the interaction of these
transcription factor proteins with the RNA polymerase complex in the regulatory region
of the gene. At the bacterial level the mechanism is the following. 
The code segment of the DNA chain is read by RNA polymerase complex (RNAp) 
thus producing the RNA messenger acid (mRNA). This one goes to the ribozomal machine which produces the 
protein according to the codon distribution in the gene segment. This protein will be at its turn the transcription factor
for a new transcriptional process.
In order to build a mathematical model of this process, one must first describe the
binding of the RNA polymerase molecule to the DNA promoter, namely a region which is the
beginning of the encoded string. In a thermodynamical description \cite{6}, the promoter activity is
proportional to the equilibrium probability g of the binding of the RNA polymerase to the core promoter. 
In the case of the 
{\it simplest} processes, namely {\it simple activation or suppression}, the
dependence of g on the cellular TF concentrations (which we shall denote by p) is described
by the Arrhenius form \cite{7}
$$g_A(p)=\frac{1 + \omega_A p/K_A}{1 + p/K_A}$$ for activation 
and
$$g_R(p) =\frac{1}{1 + p/K_R}+ \Lambda$$ for suppression
where $\omega_A=\exp(-\Delta G_{A-P}/RT)$ is the Boltzmann weight of the activator-RNAp interaction, 
$K_A$ dissociation constant between the protein and respective operator sequence in the regulatory region and $\Lambda$ is the effect of 
promoter leakage in repression. 
Also for the ribozomal activity the function describing this will have a linear form as
$$f(m)=\nu m-\mu$$
where $\nu$ is the protein synthesis rate at full activation which can have a large span of values (0-100nM/min) and $\mu$ is related to the 
fact that a certain amount of mRNA is not coded \cite{8}. 

From these basic ingredients, we can write the dynamical equations for one gene. Two
steps can be distinguished. First the RNA polymerase produces RNA-messenger acid $m$
$$dm/dt = g_{A,R}(p)-\lambda_m m$$
where $\lambda_m^{-1}$ is the mRNA half-life which is around 5 min. 
Next, the RNA-messenger acid goes to a ribosomal machine and TF proteins are produced
according to the equation:
$$dp/dt =f(m)-\gamma p= \nu m-\mu-\gamma p$$
where $\gamma^{-1}$ is the protein half-life which is ten times higher than mRNA one. 

Since the kinetics of RNA messenger production are rapid compared to those of the TF
proteins, it is not unreasonable to make a steady-state assumption for the reaction leading to
their production, and thus we have $m = g(p)/\lambda_m$. In this case the 
equation for the transcription factor production associated to a single gene is given by:
\begin{equation}
\frac{dp}{dt}=\frac{\nu}{\lambda_m}g_{A,R}(p)-\mu-\gamma p \label{prot} 
\end{equation}

In a more general context a detailed analysis of the 
interaction of RNA polymerase with the DNA chain shows that 
the binding probability has a more general form as a "stiffer" sigmoid namely,
\begin{equation}
g_{A,R}(p)=\frac{\alpha+\beta p^{\sigma}}{1+p^{\sigma}} \label{prom}
\end{equation} 
where 
$\alpha$ and $\beta$ are strictly positive numbers (for instance $\beta/\alpha$ in the activator case is proportional to the Boltzmann 
weight $\omega_A$ and $p$ is normalised by dividing it to the dissociation constant 
$p\to p/K_A$). 
Activation occurs for $\alpha<\beta$ whilst repression for $\alpha>\beta$. Experimentally in the activator case $10<\beta/\alpha<100$ and in the repressor case is slighly bigger \cite{1}. The exponent 
$\sigma$ is the Hill coefficient which can take only nonegative values. 
Biologically it represents the cooperativity in the promoter activity and is 
related to the number of operator-bound transcription factors interacting with RNAp.

In this paper we shall consider a chain of genes where each gene is in interaction with two
others, the effect of which can be either activating or suppressing. From the one-gene model
we presented above, we can generalise the equation (\ref{prot}) for any gene labeled with $n$ write as:
\begin{equation}
\frac{dp_n}{dt} = g_A(p_{n+1}) + g_R(p_{n-1}) - \gamma(n) p_{n}
\end{equation}
where $g(p_n)$ is given by equation (\ref{prom}) assuming of course that all genes 
have the same promoter 
activity and neglect the self-activation of the gene $n$.
The above equation can be in some sense misleading because we simply {\it add} the 
functions $g_A(p_{n+1})$ and $g_R(p_{n-1})$. In reality the problem is much more 
intricate since the possible interaction
between the transcription factors $p_{n+1}$ and $p_{n-1}$ can modify 
drastically the promoter activity 
function \cite{6} and a more elaborate model must take it into account . 
These interaction may appear not only because of the overlapping 
between them but also from the possible DNA looping \cite{9}.
In our model we assume that there is no interaction between transcription factors acting 
independently and thus the sum of the functions is a good approximation.

Here we are going to consider a gene network with activation-repression
interaction between the genes for general $\sigma$.
Accordingly the chain equation is given by,
$$\frac{dp_n}{dt}=\frac{\alpha+\beta p_{n+1}^{\sigma}}{1+p_{n+1}^{\sigma}}+
\frac{\beta+\alpha p_{n-1}^{\sigma}}{1+p_{n-1}^{\sigma}}-\mu-\gamma(n) p_n$$
Here we have a strongly nonlinear differential-difference equation which gives the
distribution in time of the transcription factors for all the genes
In order to put it into a more tractable form we are going to use the following substitution
(valid since $p_n(t)$ is always positive) 
$$p_n=e^{2\phi_n/\sigma}$$ and consider that
$\beta=\alpha+L$ where $L$ is a positive number. 
Then the equation will take the form:
\begin{widetext}
$$\frac{d}{dt}\left(e^{\frac{2\phi_n}{\sigma}}\right)=2\alpha-\mu+
L\left( \frac{e^{2\phi_{n+1}}}{1+e^{2\phi_{n+1}}}
+\frac{1}{1+e^{2\phi_{n-1}}} \right)-\gamma(n) e^{\frac{2\phi_n}{\sigma}}
$$
\end{widetext}
Using the fact that 
$$\frac{1}{2}(1+\tanh x)=\frac{e^{2x}}{1+e^{2x}}\quad\frac{1}{2}(1-\tanh x)=\frac{1}{1+e^{2x}}$$
then after time rescaling with $L$ one gets:

\begin{widetext}
$$\frac{d}{dt}\left(e^{\frac{2\phi_n}{\sigma}}\right)=
(\tanh\phi_{n+1}-\tanh\phi_{n-1})+(2\alpha-\mu+L)/L-(\gamma(n)/L) e^{\frac{2\phi_n}{\sigma}}$$

\end{widetext}
Since $\alpha$ and $\gamma(n)$ are small the above equation can be seen as being:
$$\left(e^{\frac{2}{\sigma}\phi_n}\right)_t=\tanh\phi_{n+1}-\tanh\phi_{n-1}+perturbation$$

Now in order to see the structure of solutions we are going to analyse the above equation without perturbation for different $\sigma$'s. 
For $\sigma=1$ putting
$$\phi_{n}=\frac{1}{2}\log(\frac{1}{u_{n}}-1),\quad 0<u_n(t)<1$$
one gets the celebrated integrable differential-difference Koretweg de Vries (KdV) equation
$$\dot u_n=u_n^2(u_{n+1}-u_{n-1})$$
It is known that it admits multisoliton solution for arbitrary number of solitons and any value of wave parameters. Here we have to analyse if this type of solution exists in the interval $(0,1)$.
The simplest way of computing soliton solution is by means of Hirota bilinear formalism \cite{10}. 
Considering the solution $u_n=F_{n+1}F_{n-1}/F_{n}^2$ the equation is turned into a bilinear one
$${\bf D}_{t}F_n\bullet F_{n+1}-F_{n-1}F_{n+2}+F_{n}F_{n+1}=0$$
where the first term is written using the Hirota bilinear operator (${\bf D}_{t}a\bullet b=\dot a b-a\dot b$).
The 1 soliton solution is given by $F_n=1+\exp(kn+\omega t)$ with $\omega=2\sinh k$ and $k$ is the free wave parameter. Immediately one can see that $u_n=F_{n+1}F_{n-1}/F_{n}^2>1$ for every $k$ so the soliton
is not in the interval. If we try a dark type soliton (a {\it hole} shape instead of a pulse shape having value $w_0$ for $n$ very big) namely $u_n=w_0-G_n/F_n$ the the equation is turned into a more complicated bilinear one,
$${\bf D}_t G_n\bullet F_n=G_{n+1}F_{n-1}-G_{n-1}F_{n+1}$$
$$F_{n+1}F_{n-1}=F_{n}^2-(2/w_0)G_{n}F_{n}+(1/w_0^2)G_{n}^2$$
The 1-soliton solution is $G_n=\exp(kn+\omega t), F_{n}=1+b\exp(kn+\omega t)+c\exp(2kn+2\omega t)$
where $\omega=2\sinh k$, $b=-2e^k/w_0(e^k-1)^2$ and $c=e^{2k}/w_0^2(e^{2k}-1)^2+4e^{3k}/w_0^2(e^k+1)^2(e^k-1)^4$
Here, because $F_n$ at the denominator have negative terms it is possible that solution may explode in finite time. This means that at some gene the protein production increase suddenly to the saturation.
The values for $k$ such that $u_{n}$ be in (0,1) are exactly in this exploding region. Accordingly, for
$\sigma=1$ there is a dark soliton-like proteomic wave which will "fire" a specific gene due to the singular character and make it work at saturation.

In the case $\sigma>>1$ making an expansion in the left exponential and rescaling time we obtain:
$$ \frac{d\phi_n}{dt}=\tanh\phi_{n+1}-\tanh\phi_{n-1}$$
and with the substitution $\phi_{n}=\tanh^{-1} w_n$ with $-1<w_n<1$
$$\dot w_n=(1-w_n^2)(w_{n+1}-w_{n-1})$$
which is the {\it defocusing} differential-discrete modified Korteweg de Vries (mKdV) equation.
This equation which has been analysed in detail in \cite{11} admits only dark-type soliton as localised solutions.

Assuming that the positive nonzero boundary is $0<w_0<1$ the 1 dark soliton has the following 
rather complicated form:
$$w_n=\frac{G_{n+1}F_{n-1}(1+w_0)-G_{n-1}F_{n+1}(1-w_0)}{G_{n+1}F_{n-1}(1+w_0)+G_{n-1}F_{n+1}(1-w_0)}$$
where
$$G_n(t)=1+a e^{kn-\omega t}, F_n(t)=1+e^{kn-\omega t},$$ 
and $\omega=2(1-w_0^2)\sinh k$, 
$$a=\frac{(1+w_0)\exp(-k)+(1-w_0)\exp(k)-2}{(1+w_0)\exp(k)+(1-w_0)\exp(-k)-2}$$
which gets wider and wider as $k$ increase. They are in the domain $(-1,1)$ and the exploding region is avoided provided $|k|<\log((1+w_0)/(1-w_0))$ . Moreover the integrable character of the equation exhibits 
general dark multisoliton solutions which at the interaction flip polarity \cite{11}
Multisoliton solution are constructed as usual bilinear Hirota formalism does.
Biologically, these solitons can be seen as a propagation of proteomic signal  for some genes in {\it off} or lower 
"gene expression" state in a background of proteins for genes in {\it on} or higher "gene expression" state. Of course the {\it on}-{\it off} picture is approximative since the level of gene expression
 is given by the soliton amplitude (depth in the dark case) in various points be more complicated. The interesting 
fact is that the bigger is the amplitude (i.e. the closer is $k$ to $\log((1+w_0)/1-w_0))$) 
the wider is the {\it hole} in the soliton shape. Moreover when 
two different amplitude solitons collide the deeper one swallows the other and 
during the interaction the last one is turned into a {\it bright} one (reversed polarity), 
namely in the {\it off} region appear some genes in the
state {\it on}. This picture occurs for an arbitrary number of dark solitons in interaction. Moreover 
$k$ is completely free so is the amplitude i.e. gene expression. The presence of perturbation will vary the amplitude
and wave numbers in a decreasing way.

Biologically ,
these solitons are propagating in "space of genes" since $n$ is the index of the 
gene and it can be in any position subjected only to the interaction with specific 
transcription factors generated by the genes $n+1$ and $n-1$. In addition
they represent the transcription factors distribution in time on the genes.
In bacteria usually $\sigma$ is not very big, so apparently 
the condition for obtaining the above discrete mKdV is missing.
In any case had we started with the following promoter-activity function
$$g_A(p)=\alpha+L \tanh(p)\quad g_R(p)=\alpha+L-L\tanh(p)$$
(having a sigmoidal shape as well) we would have obtained 
$$ \frac{d p_n}{dt}=\tanh p_{n+1}-\tanh p_{n-1}+(2\alpha-\mu+L)/L+\gamma(n)/L p_n$$
which can be transformed immediately in the perturbed defocusing mKdV and  with the same 
substitution but confined into a smaller domain $0<w_n<1$. This in 
turn decrease the domain for $k$ in the soliton solution.
Accordingly, the distribuition of transcription factors is given by the solution of a perturbed 
nonlinear semidiscrete evolution equation which for one operator site is KdV (in the nonsolitonic sector)
and for multi-operator sites is the defocusing mKdV (in the dark-soliton sector).

Now many important things arise from this model. First of all the parameters involved in our equations are tunable
($\alpha, \beta, \nu$) \cite{6}. As an outcome one can in principle control de genomic signal propagation. 
Moreover the integrable character 
of the equation shows that the results are valid for a large domains of wave numbers (k) in the solitonic solutions.
So, in principle one can start with many possible initial conditions and to expect a nice dynamics. Of course
from the experimental point of view this is extremely difficult but this is exactly what 
one wants from a model - to exhibit
a rather stable dynamics and to have trains of proteomic signals from various localised initial conditions. 
Even though nowadays usually the experimentators work with a few 
number of genes it is to be expected that in the future will be possible to construct big modules with big number of 
genes in interaction. Our model will serve as a possible robust transmission network similar to the solitonic fiber 
optical devices.

Of course many open questions remain. First of all we did 
not consider the stochastic effects in the production of 
mRNA which is extremely important.
The proper execution of the genetic program relies on
faithful signal propagation from one gene to
the next. This process may be hindered by
stochastic fluctuations arising from gene expression,
because some of the components in
the network may be  present at low numbers, which
makes fluctuations in concentrations unavoidable \cite{12}.
However, how expression fluctuations
propagate from one gene to the next
is largely unknown. But at least a model should include the 
stochastic term in the equation.
Another aspect which we neglected is the fact that equation for 
mRNA production is in fact a differential-delay since in the 
promoter activity function everything should be taken at delayed time. 
This delay is a very important ingredient in bistability or 
oscillation analysis and of course should be included in more elaborate model.

\end{document}